\documentclass[aps,preprint,amsmath,amssymb]{revtex4}
\usepackage{graphicx}
\begin{document}
\title{Analysis of excited neutrinos at the CLIC}

\author{M. K\"{o}ksal}
\email[]{mkoksal@cumhuriyet.edu.tr} \affiliation{Department of
Physics, Cumhuriyet University, 58140, Sivas, Turkey}

\begin{abstract}

We analyze the single and pair production of excited neutrinos at the CLIC through the processes $e^{+}e^{-}\rightarrow \nu^{*} \bar{\nu}$,
$e^{+}e^{-} \rightarrow e^{+}\gamma^{*} e^{-} \rightarrow e^{+} \nu^{*} W^{-}$, and $e^{+}e^{-} \rightarrow e^{+}\gamma^{*} \gamma^{*} e^{-} \rightarrow e^{+} \nu^{*} \bar{\nu}^{*} e^{-}$ ($\gamma^{*}$ is the Weizsacker-Williams photon). We examine comprehensively the excited neutrino signal and corresponding backgrounds  to obtain limits on excited neutrino mass for various values of the integrated luminosity and center-of-mass energy. We show that the obtained  bounds are more restricted than current experimental bounds.
\end{abstract}

\maketitle

\section{Introduction}

The Standard Model (SM) has been considerably successful in describing the physics of electroweak interactions. In spite of this, some questions
still remain unanswered, particularly, the number of fermion generations and their complex pattern of mixing angles and masses have not been explained by
the SM.  These problems are considered to have been answered by composite models \cite{1,2}. In these models, the known leptons and quarks
can be regarded as the ground state to a rich spectrum of excited states. Thus, charged ($e^{*}$, $\mu^{*}$ and $\tau^{*}$) and neutral ($\nu_{e}^{*}$, $\nu_{\mu}^{*}$ and $\nu_{\tau}^{*}$) excited leptons would be an incontrovertible proof for compositeness  \cite{3}.

So far, any signal for excited neutrinos were not observed at the colliders. The mass limits of excited neutrinos were found to be $m_{*}>190$ GeV from its single production, and $m_{*}>102.6$ GeV from its pair production at the LEP \cite{4,5}. An excited neutrino with a mass less than $213$ GeV by H1 collaboration at the HERA has excluded assuming $f/\Lambda=1/m_{*}$ and $f=f'$ \cite{6}. In the literature, the excited neutrinos have been experimentally or theoretically examined at the HERA \cite{HERA,HERA1,HERA2,HERA3,HERA4}, LEP \cite{LEP,LEP1,LEP2}, and LHC \cite{7,8,9}.

The interaction between excited fermions, gauge bosons and ordinary fermions is described by \cite{lag1,lag2,lag3,lag4,lag5,lag6}
\begin{eqnarray}
\textit{L}=\frac{1}{2\Lambda}\bar{\textit{l}}_{R}^{*} \sigma^{\mu\nu}[gf \frac{\vec{\tau}}{2} \cdot \vec{W}_{\mu\nu}+g' f' \frac{Y}{2} B_{\mu\nu}]\textit{l}_{L}+h.c.
\end{eqnarray}
where $\Lambda$ is the scale of the new physics responsible for the existence of excited neutrinos, $\sigma^{\mu\nu}=i\{\gamma^{\mu} \gamma^{\nu}-\gamma^{\nu} \gamma^{\mu}\}/2$ with $\gamma^{\mu}$ being the Dirac matrices, $\vec{\tau}$ represents the Pauli matrices, $Y$ is the weak hypercharge, $f$ and $f'$ are the coupling parameters associated with the corresponding gauge groups, $\vec{W}_{\mu\nu}$ and $B_{\mu\nu}$ are the field strength tensors of the $SU(2)$ and $U(1)$ with the coupling constants $g$ and $g'$, respectively.

The excited neutrino-lepton-gauge bosons vertices through above effective Lagrangian  can be obtained as follows

\begin{eqnarray}
\Theta_{\alpha}^{\nu^{*}\nu \gamma}=\frac{g_{e}(f-f')I_{3}}{2\Lambda}q^{\beta}\sigma_{\alpha\beta}(1-\gamma_{5})
\end{eqnarray}

\begin{eqnarray}
\Theta_{\alpha}^{\nu^{*}eW}=\frac{g_{e}f}{2\sqrt{2}\Lambda \,\textmd{sin}\,\theta_{W}}q^{\beta}\sigma_{\alpha\beta}(1-\gamma_{5})
\end{eqnarray}

\begin{eqnarray}
\Theta_{\alpha}^{\nu^{*}\nu Z}=\frac{g_{e}(f \, \textmd{cot}\,\theta_{W}+f' \, \textmd{tan}\,\theta_{W})I_{3}}{2\Lambda}q^{\beta}\sigma_{\alpha\beta}(1-\gamma_{5})
\end{eqnarray}

An excited neutrino decays into a SM lepton and a gauge boson.
Therefore, excited neutrinos can have a total of three decay modes: charged weak decay $\nu^{*}\rightarrow eW$, neutral weak decay $\nu^{*}\rightarrow \nu Z $ and radiative decay $\nu^{*}\rightarrow \nu \gamma$. In addition, ignoring ordinary lepton masses the
decay widths for the various electroweak decay channels of the excited neutrino
are given by

\begin{eqnarray}
\Gamma(\nu^{*}\rightarrow \nu \gamma)= \frac{1}{4}\frac{g_{e}^{2}}{4\pi}  f_{\gamma}^{2}\frac{m_{*}^{3}}{\Lambda^{2}}
\end{eqnarray}

\begin{eqnarray}
\Gamma(\nu^{*}\rightarrow \nu V)= \frac{1}{8}\frac{g_{V}^{2}}{4\pi} f_{V}^{2}\frac{m_{*}^{3}}{\Lambda^{2}}(1-\frac{m_{V}^{2}}{m_{*}^{2}})^{2}(1+\frac{m_{V}^{2}}{2m_{*}^{2}})  \,\,\,\,\,\,\,  (V=W,Z)
\end{eqnarray}
where $f_{\gamma}=(f-f')/2$, $f_{Z}=(f \textmd{cot}\,\theta_{W}+f' \textmd{tan}\,\theta_{W})/2$,
$f_{W}=f/\sqrt{2}\,\textmd{sin}\,\theta_{W}$. $g_{e}=\sqrt{4\pi\alpha}$, $g_{W}=g_{e}/\textmd{sin}\,\theta_{W}$ and $g_{Z}=g_{e}/\textmd{sin}\,\theta_{W} \, \textmd{cos}\,\theta_{W}$ are the electroweak coupling constants.
The $f$ and $f'$ couplings are specified by the composite dynamics, and they are not usually equal to each other in the case of the process of single
excited neutrino production. Also, the $\nu^{*} \nu \gamma$ coupling is proportional to $f-f'$ term. If $f$ and $f'$ are not equal to each other, then radiative decay is allowed for an excited neutrino. Excited electrons could contribute to the anomalous magnetic moment of electrons. The possible contributions to the anomalous magnetic moment
of electron have been examined by Ref. \cite{anol}. Using results from these theoretical calculations
and measurements on anomalous magnetic
moments of the electron, $f$ and $f'$ parameters can be constrained under the assumptions $f=f'$ with $\Lambda=m_{*}$ \cite{anol1}. However, they are generally used as $f=f'=1$ or $f=-f'=1$ with $\Lambda=m_{*}$ in the literature.
The branching ratios and total decay widths depending on the mass of excited neutrinos are given in Tables I and II. At $f=f'=1$ the branching ratios for large values of excited neutrino mass reach $61\%$ for charged current decays, and $39\%$ for neutral current decays. In addition, for the case $f=-f'=1$ the photon channel does not vanish and the branchings for the charged, radiative and neutral current channels are obtained as $61\%$, $28\%$, and $11\%$. Hence, we assume that excited neutrinos only decay to the charged current channel, since this channel is dominant.

The Large Hadron Collider (LHC) is expected to answer some of the fundamental open questions in particle physics. On the other hand, it may not provide precision measurements due to the typical characteristic of a hadron machine.
A TeV scale  linear $e^{+}e^{-}$ collider with high-luminosity is the best option to complement and to extend the
LHC physics program. The Compact Linear Collider (CLIC) is a planned linear collider concept based on normal conducting accelerating cavities and two beam acceleration \cite{17}. It is proposed to carry out collisions at energies from $0.5$ to $3$ TeV. It has been foreseen that this collider will be performed in several research regions \cite{18}. These research regions are given in Table III. The first stage of the CLIC will allow the obtainment of high precision measurements of various observables of the SM Higgs boson, top and gauge sectors \cite{18}. The second stage will give access to the discovery of new physics beyond the SM. Additional Higgs features such as the Higgs self-coupling and the top-Yukawa coupling and rare Higgs decay modes will be analyzed at this stage \cite{19}. The final stage of CLIC operating at an energy of $3$ TeV is expected to provide the most precise measurements of the SM, and to directly examine the pair-production of new heavy particles of mass up to $1.5$ TeV \cite{18}.

A new possibility anticipated for the linear colliders is to operate
this machine as $e \gamma$ and $\gamma \gamma$
colliders \cite{35,36}. This can be realized by converting
the incoming leptons into an intense beam of high-energy photons.
The linear colliders also provide the opportunity to examine the $\gamma ^{*} \gamma^{*}$ and $e \gamma^{*}$ collisions with quasi real photons \cite{39,40,41,42}. $\gamma ^{*} \gamma^{*}$ and $e \gamma^{*}$ collisions can occur spontaneously with respect to $\gamma \gamma$ and $e \gamma$ collisions. For this reason, the investigation of new physics beyond the SM with $\gamma ^{*} \gamma^{*}$ and $e \gamma^{*}$ collisions is more realistic than $\gamma \gamma$ and $e \gamma$ collisions occurring via the laser backscattering
procedure. $\gamma^{*}$ photons are emitted from one of the
$e^{-}$ or $e^{+}$ beams and defined by the Weizsacker-Williams approximation (WWA). In the WWA, the virtuality of the photons which carry
a small transverse momentum is very low. Hence, the photons emitted from
$e^{-}$ or $e^{+}$ beams are generally scattered with very
small angles from their trajectory along
the beam path. This approximation has many advantages.
It allows the ability to obtain crude numerical estimations through simple formulas \cite{WWA}. Also, it may
substantially ease the experimental analysis because it enables one to obtain directly an approximate
cross section for $\gamma^{*} \gamma^{*}\rightarrow X $ process through the analysis of the process $e^{-}e^{+}\rightarrow e^{-}Xe^{+}$
\cite{WWA}. Finally, these processes have a very clean experimental environment, since they have no
interference with weak and strong interactions.

Photon-induced processes have been experimentally or theoretically investigated in the framework of the WWA at the LEP, Tevatron, and LHC \cite{399,400,401,402,43,44,45,q1,q2,q3,q4,q5,q6,q7,kok,kok2,kok22,kok1,q8,q9,q10,q11,q12,q13}. Furthermore, new physics studies are performed by making use of WWA at the CLIC in literature \cite{451,46,47,555}. In Ref. \cite{451}, the production and signatures of doubly
charged Higgs bosons in the process $\gamma^{*} e \rightarrow H^{--}E^{+}$ have been investigated.
However, Ref. \cite{46} examined the electromagnetic moments of the tau lepton in the process
$e^{+}e^{-} \rightarrow e^{+}\gamma^{*} \gamma^{*}e^{-} \rightarrow e^{+}\tau^{*} \tau^{*} e^{-}$.
The anomalous $tq\gamma$ interactions via the process $e^{+}e^{-} \rightarrow e^{+}\gamma^{*} \gamma^{*}e^{-} \rightarrow e^{+}t \bar{q} e^{-}$ have been analyzed by Ref. \cite{47}. Finally, Ref. \cite{555} has studied the signals for composite vector leptoquarks through $\gamma^{*} e $ and $\gamma^{*} \gamma^{*}$ collisions.

Ref. \cite{8} has shown that the excited neutrinos via the process $pp\rightarrow e\nu^{*} \rightarrow e \nu \gamma$ at LHC could be accessible up to a mass of
1.5 TeV, assuming an integrated luminosity of $L=300$ fb$^{-1}$ and $f=-f'=1$. However, single production of excited neutrinos in $ep$ colliders through $eq\rightarrow \nu^{*} q'\rightarrow W^{+} e q'$ supposing $f=f'=1$ has been investigated by Ref. \cite{9}. They found that excited neutrinos can be discovered up to the mass of $1300$ GeV. According to Ref. \cite{9}, at the same center-of-mass energy, the most ideal collider for investigating excited neutrinos between $ee$, $ep$ and $pp$ colliders is an $ee$ collider. For this reason, the CLIC provides an excellent opportunity to examine the signal of the excited neutrino.
Therefore, we investigated the single and pair production of excited neutrinos through the processes $e^{+}e^{-}\rightarrow \nu^{*} \bar{\nu}$, $e^{+}e^{-} \rightarrow e^{+}\gamma^{*} e^{-} \rightarrow e^{+} \nu^{*} W^{-} $, and $e^{+}e^{-} \rightarrow e^{+}\gamma^{*}\gamma^{*} e^{-} \rightarrow e^{+}\nu^{*} \bar{\nu}^{*} e^{-}$ at the CLIC.

\section{Single production via the process $e^{+}e^{-}\rightarrow \nu^{*} \bar{\nu}$}

The production mechanism for the single excited neutrino in $e^{+}e^{-}$ collision at the CLIC is given by the process $e^{+}e^{-}\rightarrow \nu^{*} \bar{\nu}$ as shown in Fig. $1$. This process is described by three tree-level diagrams, and its polarization summed amplitude is given as follows

\begin{eqnarray}
|M_{1}|^{2}=-\frac{4 g_{e}^{4}f_{\gamma}^{2}}{\Lambda^{2}s}(m_{*}^{4}+2t(t+s)-m_{*}^{2}(2t+s)),
\end{eqnarray}

\begin{eqnarray}
|M_{2}|^{2}=&&-\frac{g_{e}^{4}f_{Z}^{2}}{4 \Lambda^{2} \sin^{2}\theta_{W} \cos^{2}\theta_{W}  (s-m_{Z}^{2})^{2}} (m_{*}^{4} (1-c_{V})^{2} \nonumber \\
&&+2t(t+s)(1+4 c_{V}^{2})+m_{*}^{2}(1-2c_{V})^{2}(s+2t)), \nonumber \\
\end{eqnarray}

\begin{eqnarray}
|M_{3}|^{2}=-\frac{2 g_{e}^{4}f_{W}^{2}}{\Lambda^{2}\sin^{2}\theta_{W}(t-m_{W}^{2})^{2}}(t(t+s-m_{*}^{2})(s-m_{*}^{2})),
\end{eqnarray}

\begin{eqnarray}
|M_{12}|^{2}= &&\frac{2 g_{e}^{4}f_{Z}f_{\gamma}}{\Lambda^{2}\sin \theta_{W} \cos \theta_{W}(s-m_{Z}^{2})} (m_{*}^{4} (2c_{V}-1)-m_{*}^{2}(s+2t)(2c_{V}-1)+4c_{V}t(s+t)), \nonumber \\
\end{eqnarray}

\begin{eqnarray}
|M_{13}|^{2}=\frac{4\sqrt{2} g_{e}^{4}f_{W}f_{\gamma}}{\Lambda^{2}\sin \theta_{W} (t-m_{W}^{2})}(t(s+t-m_{*}^{2})),
\end{eqnarray}

\begin{eqnarray}
|M_{23}|^{2}=\frac{\sqrt{2} g_{e}^{4}f_{W}f_{Z}(1-2c_{V})}{\Lambda^{2} \cos\theta_{W} \sin^{2} \theta_{W} (s-m_{Z}^{2})(t-m_{W}^{2})}(st(t+s-m_{*}^{2})).
\end{eqnarray}
where $c_{V}=-1/2+2 \, \textmd{sin}^{2} \theta_{W}$; $\theta_{W}$ is the Weinberg angle.

In the course of all the calculations in this paper, we used the simulation program
COMPHEP-4.5.1 to calculate the cross sections of the signal and background \cite{48}. In this work, we took into account all possible cases with regard all tree-level SM model for the analyzed processes. Tree level diagrams generate major SM contributions. Possible other SM backgrounds contributing
to these processes are at the loop level and they can be ignored when compared to tree-level SM backgrounds.  In Fig. $2$, the total cross section of the process $e^{+}e^{-}\rightarrow \nu^{*} \bar{\nu}\rightarrow e^{-} W^{+} \bar{\nu}$ as a function of the excited neutrino mass for various values of coupling parameters and center-of-mass energies is given. Actually, since excited neutrinos are not in the SM, it is impossible to examine this process. However, if excited neutrinos exists, the SM cross section of analyzed processes changes with mass of excited neutrino as seen in Fig. $2$. We also find from these figures that the total cross sections of the processes decrease with a decrease in the coupling parameter and excited neutrino mass since the amplitudes are proportional to these parameters.  In Fig. $3$, the invariant mass $M_{We}$ distribution of signal for different mass values of excited neutrinos at the $\sqrt{s}=0.5,1.5$ and $3$ TeV and parameter $f=-f'=1$ for the process $e^{+}e^{-}\rightarrow \nu^{*} \bar{\nu}\rightarrow e^{-} W^{+} \bar{\nu}$ are plotted. The size of the peak increases when the center-of-mass energy increases.

\section{Single production via the process $e^{+}e^{-} \rightarrow e^{+}\gamma^{*} e^{-} \rightarrow e^{+}\nu^{*} W^{-}$}

A quasi real photon emitted from one of the $e^{-}$ or $e^{+}$ beams can interact with the
other beam and the subprocess $e^{-}\gamma^{*} \rightarrow \nu^{*} W^{-}$ can occur at the CLIC. Hence, it would be possible to examine $e \gamma^{*}$ collisions. A schematic diagram describing
this process is presented in Fig. $4$. In the existence of the effective Lagrangian in Eq. ($1$), Feynman diagrams of the subprocess $e^{-}\gamma^{*} \rightarrow \nu^{*} W^{-}$ containing anomalous $\nu^{*}\nu \gamma$ and $\nu^{*}eW$ couplings are given in Fig. $5$. The analytical expression of squared amplitudes can be given in terms of Mandelstam variables by the formula:

\begin{eqnarray}
|M_{1}|^{2}=&&\frac{g_{e}^{4}f_{W}^{2}}{s \Lambda^{2} \sin^{2}\theta_{W}}(-m_{W}^{4}+(s+t)m_{W}^{2}+2 m_{*}^{2}(t+m_{W}^{2})-2st),
\end{eqnarray}

\begin{eqnarray}
|M_{2}|^{2}=&&-\frac{g_{e}^{2} g_{W}^{2} f_{\gamma}^{2}}{4 \Lambda^{2} m_{W}^{2}  (m_{*}^{2}+m_{W}^{2}-s-t)^{2}} (t+s-m_{W}^{2}) (s m_{*}^{4} -2(m_{W}^{4}-(s+t)m_{W}^{2} \nonumber \\
&&+s(s+t))m_{*}^{2}-2 m_{W}^{2}(s+t)^{2}+s(s+t)^{2}+m_{W}^{4}(s+2t)), \nonumber \\
\end{eqnarray}

\begin{eqnarray}
|M_{3}|^{2}=&&\frac{g_{e}^{4}f_{W}^{2}}{16 \Lambda^{2} \sin^{2}\theta_{W} m_{W}^{4} (t-m_{W}^{2})^{2}}(-m_{*}^{6}(53m_{W}^{4}-10 t m_{W}^{2}+t^{2})+(-52 m_{W}^{6} \nonumber \\
&&+2(8s+19t)m_{W}^{4}-4tm_{W}^{2}(3s-7t)+2t^{2}(2s-3t))m_{*}^{4}+(20 m_{W}^{8} \nonumber \\
&&-96sm_{W}^{6}+tm_{W}^{4}(84s+11t)+2tm_{W}^{2}(2s^{2}+4ts-17t^{2})+ t^{3} \nonumber \\
&&(7t-4s))m_{*}^{2}+4t m_{W}^{2}(-t^{3}+(m_{W}^{2}+4s)t^{2}+(13m_{W}^{4} \nonumber \\
&&-24sm_{W}^{2}+4s^{2})t-5(m_{W}^{3}-2m_{W}s)^{2})),
\end{eqnarray}

\begin{eqnarray}
|M_{12}|^{2}= &&\frac{g_{e}^{3} g_{W}f_{W}f_{\gamma}}{2 \Lambda^{2}\sin \theta_{W} (s+t-m_{W}^{2}-m_{*}^{2})} (m_{*}^{4}+(m_{W}^{2}-s-2t)m_{*}^{2} \nonumber \\ &&-(m_{W}^{2}-s-t)(t+2m_{W}^{2}))
\end{eqnarray}

\begin{eqnarray}
|M_{13}|^{2}=&&-\frac{g_{e}^{4}f_{W}^{2}}{2 \Lambda^{2}m_{W}^{2}\sin^{2} \theta_{W} s(t-m_{W}^{2})}(6m_{W}^{2}m_{*}^{6}-(4m_{W}^{4}+(9s+4t) \nonumber \\
&&m_{W}^{2}+st)m_{*}^{4}+(-2m_{W}^{6}-7sm_{W}^{4}+4s^{2}m_{W}^{2}+(s-2m_{W}^{2})t^{2} \nonumber \\
&&+2t(s+m_{W}^{2})^{2})m_{*}^{2}+m_{W}^{2}t(2m_{W}^{4}+(2t-3s)m_{W}^{2}+s(t-4s))) \nonumber \\,
\end{eqnarray}

\begin{eqnarray}
|M_{23}|^{2}=&&\frac{g_{e}^{3}g_{W}f_{\gamma}f_{W}}{4 \Lambda^{2}m_{W}^{2}\sin \theta_{W}(t-m_{W}^{2})(s+t-m_{*}^{2}-m_{W}^{2})}((m_{W}^{2}-t)m_{*}^{6} \nonumber \\ &&+(8m_{W}^{2}-7s-3t)(m_{W}^{2}-t)m_{*}^{4}+(-5m_{W}^{6}+(11s+t)m_{W}^{4} \nonumber \\
&&+(-4s^{2}+5ts+7t^{2})m_{W}^{2}-t(6s^{2}+8ts+3t^{2}))m_{*}^{2} \nonumber \\
&&+t(m_{W}^{2}-s-t)(5m_{W}^{4}-4m_{W}^{2}(s+t)-t^{2})) \nonumber \\
\end{eqnarray}

If radiative decay of the excited neutrino vanishes ($f=f'$), then $\nu^{*}eW$ coupling can be isolated through the subprocess $e^{-}\gamma^{*} \rightarrow \nu^{*} W^{-}$. For this reason, we compared the total cross sections of single excited neutrino production of the process $e^{+}e^{-} \rightarrow e^{+}\gamma^{*} e^{-} \rightarrow e^{+}\nu^{*} W^{+}\rightarrow e^{+}e^{-}W^{+}W^{-}$  as a function of excited neutrino mass at $f=-f'=0.5,1$ and $f=f'=0.5,1$ values to understand the effects of $f_{\gamma}$ in an excited neutrino production mechanism in Fig. $6$. It can be seen from these figures that the total cross sections of the process at $f=-f'$ coupling parameters are greater than $f=f'$ values. The invariant mass $M_{We}$ distributions of signal for different mass values of excited neutrinos are given in Fig. $7$, assuming $f=-f'=1$.

\section{Pair production via the process $e^{+}e^{-} \rightarrow e^{+}\gamma^{*}\gamma^{*} e^{-} \rightarrow e^{+}\nu^{*} \bar{\nu}^{*} e^{-}$}

The almost real photons emitted from
both $e^{-}$ and $e^{+}$ beams interact with each other, and the subprocess $\gamma^{*}\gamma^{*}\rightarrow \nu^{*} \bar{\nu}^{*}$ is produced as given by Fig. $8$. The subprocess $\gamma^{*}\gamma^{*}\rightarrow \nu^{*} \bar{\nu}^{*}$ consists of t and u channel Feynman diagrams, as shown in Fig. $9$. In case of effective interaction $\nu^{*}\nu \gamma$, the polarization summed amplitude square for the subprocess is given by

\begin{eqnarray}
|M_{1}|^{2}=&&-\frac{16 g_{e}^{2}f_{\gamma}^{2}}{s \Lambda^{2}}(m_{*}^{2}-t)^{2}(m_{*}^{4}-2t m_{*}^{2}+t(s+t)),
\end{eqnarray}

\begin{eqnarray}
|M_{2}|^{2}=|M_{1}|^{2}(t \leftrightarrow u)
\end{eqnarray}

\begin{eqnarray}
|M_{12}|^{2}=0.
\end{eqnarray}

We will not review the process $e^{+}e^{-} \rightarrow e^{+}\gamma^{*} \gamma^{*} e^{-} \rightarrow e^{+} \nu^{*} \bar{\nu}^{*} e^{-}$
at $\sqrt{s}=0.5$ TeV, since the experimental mass limit of the excited neutrino is $213$ GeV. The SM and new physics total cross sections of the process $e^{+}e^{-} \rightarrow e^{+}\gamma^{*} \gamma^{*} e^{-} \rightarrow e^{+} \nu^{*} \bar{\nu}^{*} e^{-} \rightarrow e^{+} e^{-} W^{+} W^{-}e^{+}e^{-}$ as a function of the excited neutrino mass at values of $f=-f'=0.5,1$ coupling parameters and $\sqrt{s}=1.5,3$ TeV is depicted in Fig. $10$.

\section{Numerical Results}

We need to carry out statistical analysis for a detailed examination of the excited neutrino signal.
In this study, we estimate sensitivity of the processes $e^{+}e^{-}\rightarrow \nu^{*} \bar{\nu}\rightarrow e^{-} W^{+} \bar{\nu}$,
$e^{+}e^{-} \rightarrow e^{+}\gamma^{*} e^{-} \rightarrow e^{+} \nu^{*} W^{-} \rightarrow e^{+}e^{-}W^{+}W^{-}$, and $e^{+}e^{-} \rightarrow e^{+}\gamma^{*} \gamma^{*} e^{-} \rightarrow e^{+} \nu^{*} \bar{\nu}^{*} e^{-} \rightarrow e^{+} e^{-} W^{+} W^{-}e^{+}e^{-}$ on the mass of the excited neutrino using two different statistical analysis methods. We perform a one-parameter $\chi^{2}$ analysis when the number of SM events is greater than $10$ \cite{kok2,kok22,q12}. The $\chi^{2}$ analysis is defined by

\begin{eqnarray}
\chi^{2}=\left(\frac{\sigma_{SM}-\sigma_{AN}}{\sigma_{SM}\delta_{stat}}\right)^{2}
\end{eqnarray}
where $\sigma_{AN}$ is the total cross section containing SM and new physics, $\delta_{stat}=\frac{1}{\sqrt{N_{SM}}}$ is the statistical
error: $N_{SM}$ is the number of SM events. In the second analysis, we applied a Poisson distribution, due to the number of SM events smaller than or equal to $10$. Here, the bounds on the excited neutrino mass are calculated supposing the number of observed events equal to the SM prediction. Upper bounds of the number of events $N_{up}$ at the $95\%$ C. L. can be obtained as follows \cite{ses3,ses4}

\begin{eqnarray}
\sum_{k=0}^{N_{obs}}P_{Poisson}(N_{up};k)=0.05.
\end{eqnarray}

In addition, for each process we imposed a cut on the invariant mass $M_{We}$ to suppress the SM background and to extract the excited neutrino signal. Hence, we perform the cuts $|m_{*}-m_{We}|<25$ GeV for $m_{*}=200-1200$ GeV, $|m_{*}-m_{We}|<50$ GeV for $m_{*}=1200-2000$ GeV, and $|m_{*}-m_{We}|<75$ GeV for $m_{*}=2000-2900$ GeV.

The SM event number for the process $e^{+}e^{-}\rightarrow \nu^{*} \bar{\nu}\rightarrow e^{-} W^{+} \bar{\nu}$
is calculated as follows

\begin{eqnarray}
N_{SM}=\sigma_{SM}\times L_{int} \times BR(W\rightarrow q\bar{q})
\end{eqnarray}
where $\sigma_{SM}$ is the SM cross section, $L_{int}$ is the integrated luminosity. A $W$ boson can decay into a lepton and neutrino. Then, our process consisted of two neutrinos in the final state, and this situation caused a great uncertainty. Therefore, we took into account the hadronic decay of the
$W$ boson with the branching through $W\rightarrow q\bar{q}$. Also, we applied the cuts $p_{T}^{\,e}>20 \:$ GeV and  $|\eta_{e}|<2.5$ for the electron in the final state. Here, $p_{T}$ is the transverse momentum and $|\eta|$ is the pseudorapidity.

After applying the above condition, the SM cross sections for the process $e^{+}e^{-}\rightarrow \nu^{*} \bar{\nu}\rightarrow e^{-} W^{+} \bar{\nu}$ such as at excited neutrino masses $200$ GeV and $400$ GeV are obtained as $9.92\times10^{-2}$ pb and $1.83\times10^{-1}$ pb for a center-of-mass energy of $0.5$ TeV, respectively. In addition, we have calculated the SM cross sections as $5.51\times10^{-2}$ pb and $2.64\times10^{-2}$ pb for excited neutrino masses $200$ GeV and $1200$ GeV at $\sqrt{s}=1.5$ TeV, respectively. Finally, at $\sqrt{s}=3$ TeV, we found the SM cross sections to be $1.84\times10^{-2}$ pb and $8.03\times10^{-3}$ pb for excited neutrino masses $200$ GeV and $2600$ GeV, respectively. For the $f=-f'=0.3,0.5$ and $1$ coupling parameters, $95\%$ C.L. limits for an excited neutrino mass as a function of the integrated CLIC luminosity for the process $e^{+}e^{-}\rightarrow \nu^{*} \bar{\nu}\rightarrow e^{-} W^{+} \bar{\nu}$ at different values of center-of-mass energy are presented in Fig. $11$. We can see from these figures that the excited neutrinos for the process $e^{+}e^{-}\rightarrow \nu^{*} \bar{\nu}\rightarrow e^{-} W^{+} \bar{\nu}$ can be determined up to the center-of-mass energy of three stages of the CLIC. In particular, the accessible limits for the excited neutrino mass at $f=-f'=0.3$ and $L_{int}=10$ fb$^{-1}$ is approximately obtained as $440$, $1250$, and $2600$ GeV at $\sqrt{s}=0.5,1.5$ and $3$ TeV, respectively.

On the other hand, the SM event number for the process $e^{+}e^{-} \rightarrow e^{+}\gamma^{*} e^{-} \rightarrow e^{+}\nu^{*} W^{-} \rightarrow e^{+} e^{-}W^{+} W^{-} $ is given by

\begin{eqnarray}
N_{SM}=\sigma_{SM}\times L_{int} \times BR(W\rightarrow q\bar{q})\times BR(W\rightarrow \ell \nu_{\ell}).
\end{eqnarray}
Here, we considered that one of the $W$ bosons decayed leptonically and the other hadronically for the signal. Hence, we assumed that the branching ratio of the $W$ bosons in the final state was $BR=0.145$. In addition, we performed cuts $p_{T}^{\,e}>20 \:$ GeV and  $|\eta_{\,e}|<2.5$, for the electron in the final state. The SM cross sections for the process $e^{+}e^{-} \rightarrow e^{+}\gamma^{*} e^{-} \rightarrow e^{+}\nu^{*} W^{-} \rightarrow e^{+} e^{-}W^{+} W^{-} $, for example, at excited neutrino masses $200$ GeV and $400$ GeV, we calculated as $5.97\times10^{-3}$ pb and $3.90\times10^{-4}$ pb at $\sqrt{s}=0.5$ TeV, respectively. Also, for excited neutrino masses $200$ GeV and $1200$ GeV, we obtained the SM cross sections as $1.15\times10^{-2}$ pb and $9.21\times10^{-4}$ pb at $\sqrt{s}=1.5$ TeV, respectively. Finally, at $\sqrt{s}=3$ TeV, we found the SM cross sections to be $8.35\times10^{-3}$ pb and $6.96\times10^{-4}$ pb for the same process at excited neutrino masses $200$ GeV and $2600$ GeV, respectively. In Fig. $12$, the $95\%$ C.L. limits for an excited neutrino mass
as a function of integrated CLIC luminosity for the process $e^{+}e^{-} \rightarrow e^{+}\gamma^{*} e^{-} \rightarrow e^{+}\nu^{*} W^{-} \rightarrow e^{+} e^{-}W^{+} W^{-}$ at different values of center-of-mass energy were plotted. As shown in Fig. $12$, the obtained limits for the excited neutrino mass at $f=-f'$ values are greater than the limit values derived from $f=f'$  values.

For the process $e^{+}e^{-} \rightarrow e^{+}\gamma^{*} \gamma^{*} e^{-} \rightarrow e^{+} \nu^{*} \bar{\nu}^{*} e^{-} \rightarrow e^{+} e^{-} W^{+} W^{-}e^{+}e^{-}$, the SM event number is given by

\begin{eqnarray}
N_{SM}=\sigma_{SM}\times L_{int} \times BR(W\rightarrow q\bar{q})\times BR(W\rightarrow \ell \nu_{\ell}).
\end{eqnarray}
Also, we performed cuts $p_{T}^{\,e^{+},\,e^{-}}>20 \:$ GeV and  $|\eta_{e^{+},\,e^{-}}|<2.5$ for the electron and positron in the final state. With assuming these restrictions, at $\sqrt{s}=1.5$ TeV, the SM cross sections were found to be $4.01\times10^{-6}$ pb and $1.49\times10^{-10}$ pb for the excited neutrino masses $200$ GeV and $700$ GeV, respectively. For the center-of-mass energy of $3$ TeV, we calculated the SM cross sections to be $1.03\times10^{-5}$ pb and $1.71\times10^{-9}$ pb for the excited neutrino masses $200$ GeV and $1200$ GeV, respectively. For the process $e^{+}e^{-} \rightarrow e^{+}\gamma^{*} \gamma^{*} e^{-} \rightarrow e^{+} \nu^{*} \bar{\nu}^{*} e^{-} \rightarrow e^{+} e^{-} W^{+} W^{-}e^{+}e^{-}$, the limits of excited neutrino mass
as a function of integrated luminosity at various values of center-of-mass energy are given in Fig. $13$. At $f=-f'=1$ and $L_{int}=10$ fb$^{-1}$, the accessible limits for excited neutrino masses are obtained as $265$ and $408$ GeV at center-of-mass energies of $1.5$ and $3$ TeV. Also, at the same center-of-mass energies, we found excited neutrino masses as $240$ and $373$ GeV for $f=-f'=0.5$ and $L_{int}=100$ fb$^{-1}$.

\section{Conclusions}

Despite the fact that $ep$ and $pp$ colliders have high luminosity and high energy, they do not have a very clean environment due to the
proton remnants. On the other hand, linear $e^{+}e^{-}$ colliders with TeV
scale energy and extremely high luminosity have less background than $ep$ and $pp$ colliders. For this reason, the linear colliders
can investigate excited neutrinos with a much higher precision with respect to the $ep$ and $pp$ colliders.
According to $ep$ and $pp$ colliders, at the same center-of-mass energy and coupling parameter values, the linear colliders are more likely to determine the excited neutrinos. Therefore, we have investigated the processes $e^{+}e^{-}\rightarrow \nu^{*} \bar{\nu}$,
$e^{+}e^{-} \rightarrow e^{+}\gamma^{*} e^{-} \rightarrow e^{+} \nu^{*} W^{-} $, and $e^{+}e^{-} \rightarrow e^{+}\gamma^{*} \gamma^{*} e^{-} \rightarrow e^{+} \nu^{*} \bar{\nu}^{*} e^{+}$ at the CLIC
to examine signals of the excited neutrino. We show that the best process determining the excited neutrinos from these processes at the CLIC is $e^{+}e^{-}\rightarrow \nu^{*} \bar{\nu}$. Also, excited neutrinos can be obtained up to the center-of-mass energies of this process at $f=-f'=1$. Consequently,
the CLIC provides us an excellent opportunity to probe the excited neutrinos in a very clean environment.

\pagebreak

\pagebreak

\begin{figure}
\includegraphics{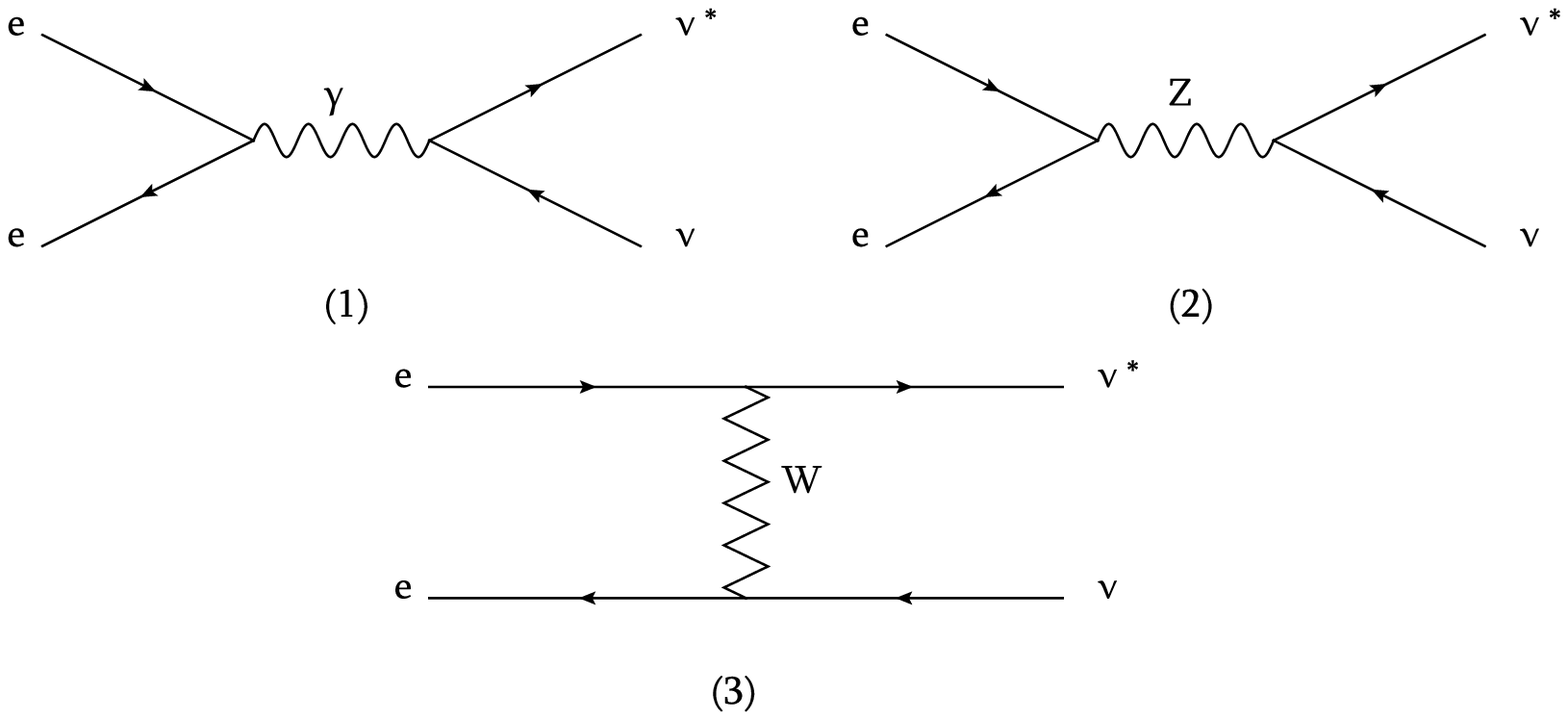}
\caption{Tree-level Feynman diagrams for the process $e^{+}e^{-}\rightarrow \nu^{*} \bar{\nu}$.
\label{fig1}}
\end{figure}

\begin{figure}
\includegraphics{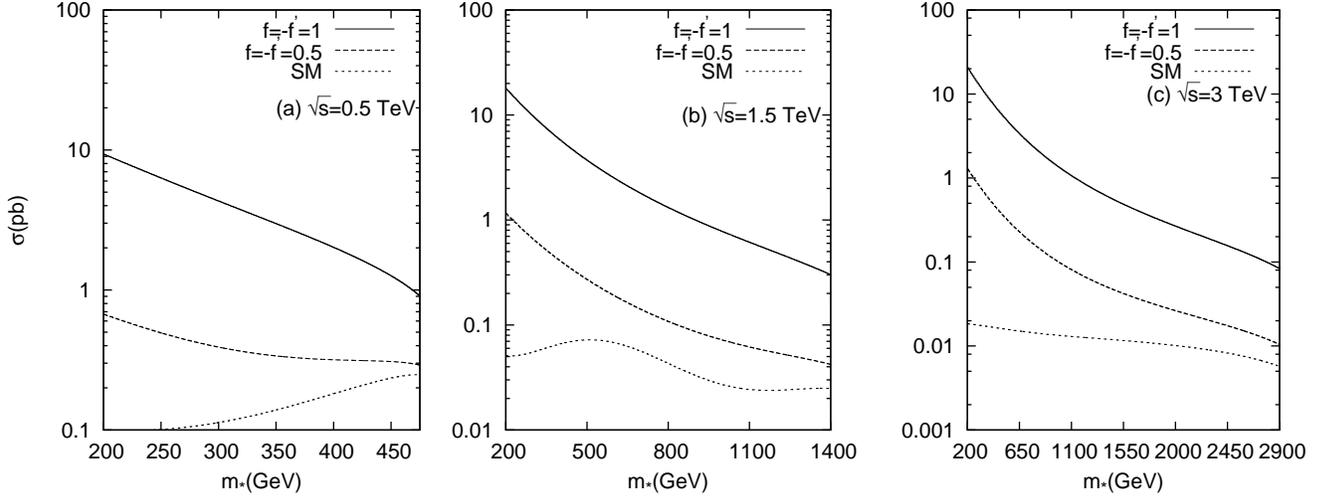}
\caption{The SM and new physics total cross sections for single excited neutrino production of the process $e^{+}e^{-}\rightarrow \nu^{*} \bar{\nu}\rightarrow e^{-} W^{+} \bar{\nu}$  as a function of excited neutrino mass at various values of coupling parameter and center-of-mass energy.
\label{fig2}}
\end{figure}

\begin{figure}
\includegraphics{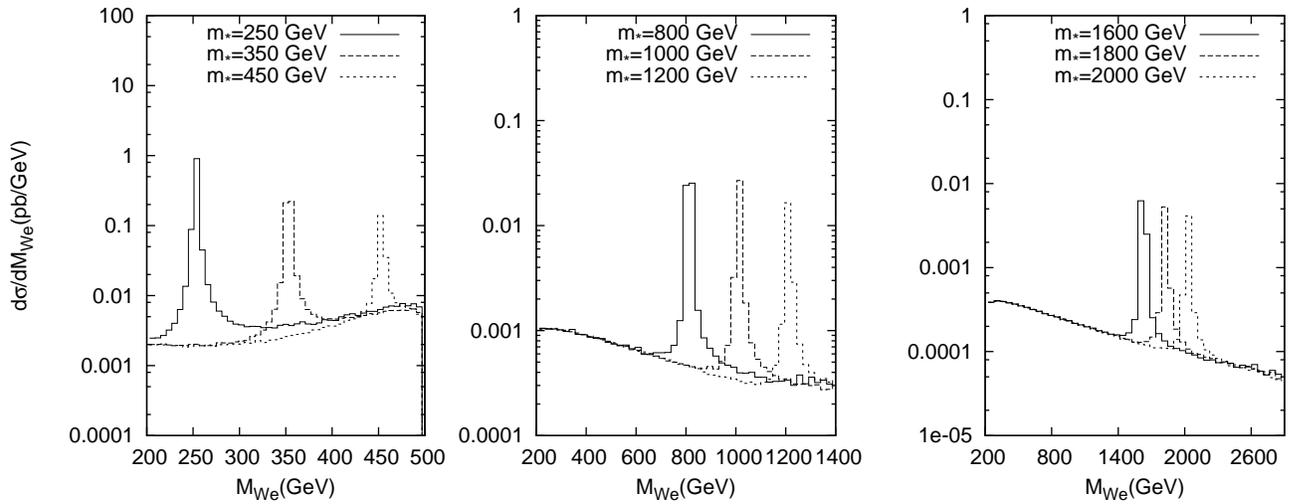}
\caption{The invariant mass $M_{We}$ distribution of signal for different mass values of excited neutrino at the
$\sqrt{s}=0.5,1.5$ and $3$ TeV for the process $e^{+}e^{-}\rightarrow \nu^{*} \bar{\nu}\rightarrow e^{-} W^{+} \bar{\nu}$.
\label{fig3}}
\end{figure}

\begin{figure}
\includegraphics{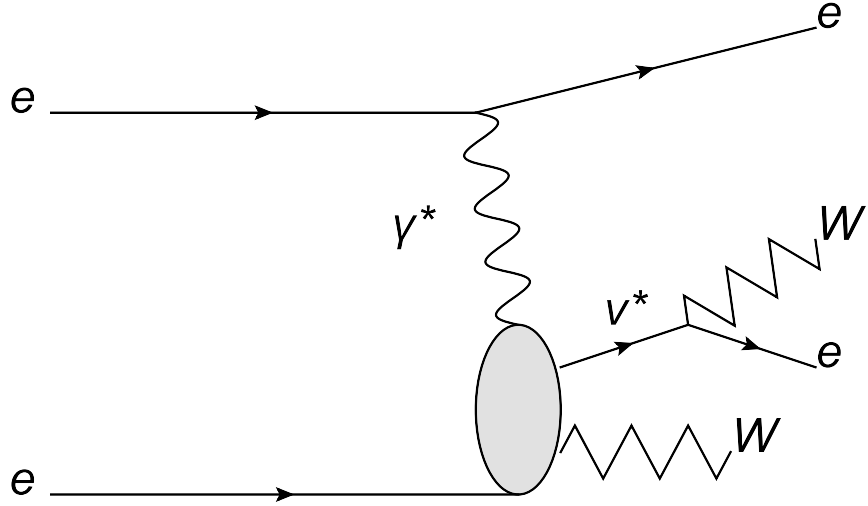}
\caption{Schematic diagram for the process $e^{+}e^{-} \rightarrow e^{+}\gamma^{*} e^{-} \rightarrow e^{+}\nu^{*} W^{-}$.
\label{fig4}}
\end{figure}

\begin{figure}
\includegraphics{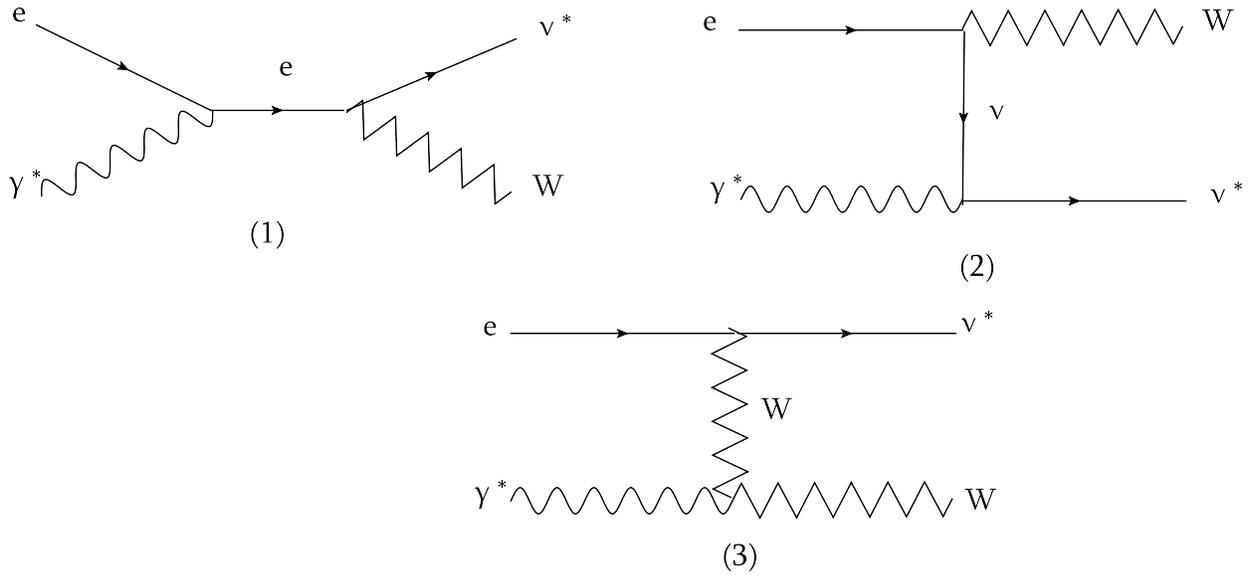}
\caption{Tree-level Feynman diagrams for the subprocess $e^{-}\gamma \rightarrow \nu^{*} W^{-}$.
\label{fig5}}
\end{figure}

\begin{figure}
\includegraphics{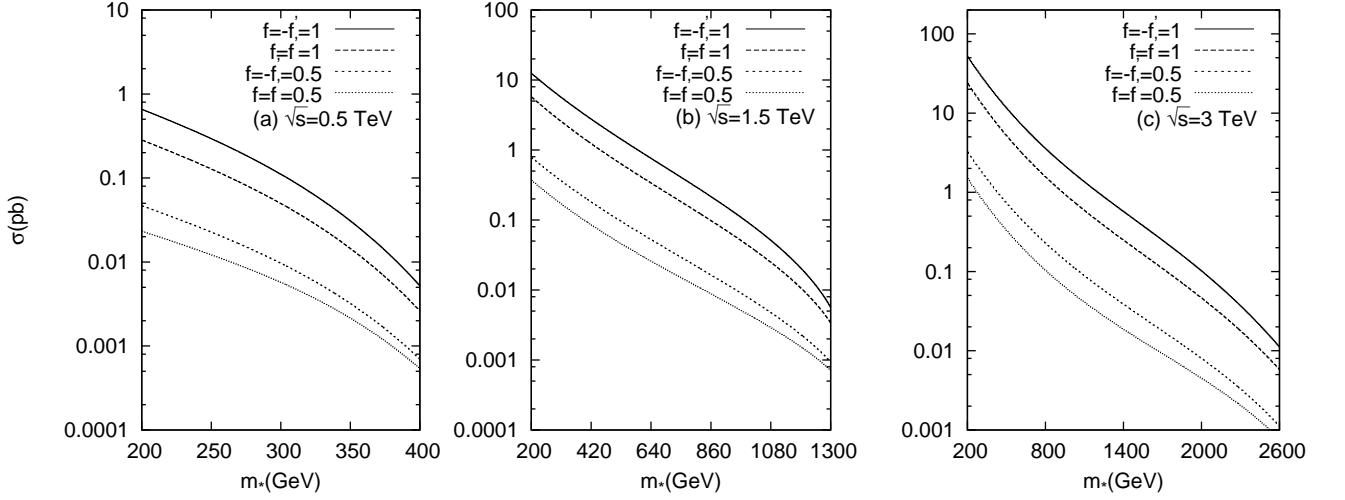}
\caption{The total cross sections for single excited neutrino production of the process $e^{+}e^{-} \rightarrow e^{+}\gamma^{*} e^{-} \rightarrow e^{+}\nu^{*} W^{-}\rightarrow e^{+}e^{-}W^{+}W^{-}$  as a function of excited neutrino mass at various values of coupling parameter and center-of-mass energy.
\label{fig6}}
\end{figure}

\begin{figure}
\includegraphics{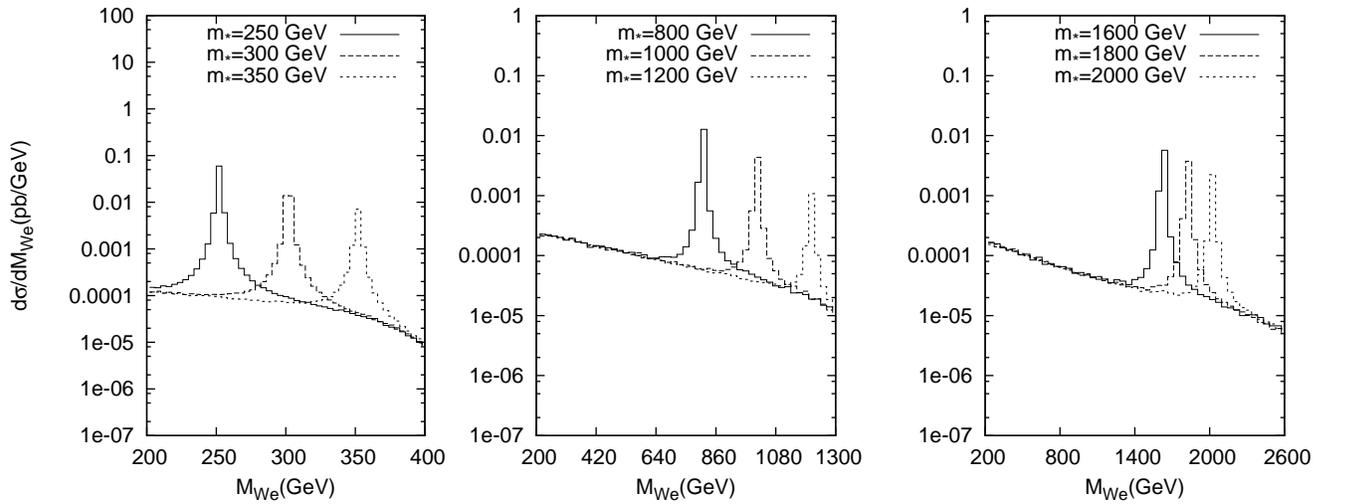}
\caption{The invariant mass $M_{We}$ distribution of signal for different mass values of excited neutrino at the
$\sqrt{s}=0.5,1.5$ and $3$ TeV for the process $e^{+}e^{-} \rightarrow e^{+}\gamma^{*} e^{-} \rightarrow e^{+}\nu^{*} W^{-}\rightarrow e^{+}e^{-}W^{+}W^{-}$.
\label{fig7}}
\end{figure}

\begin{figure}
\includegraphics{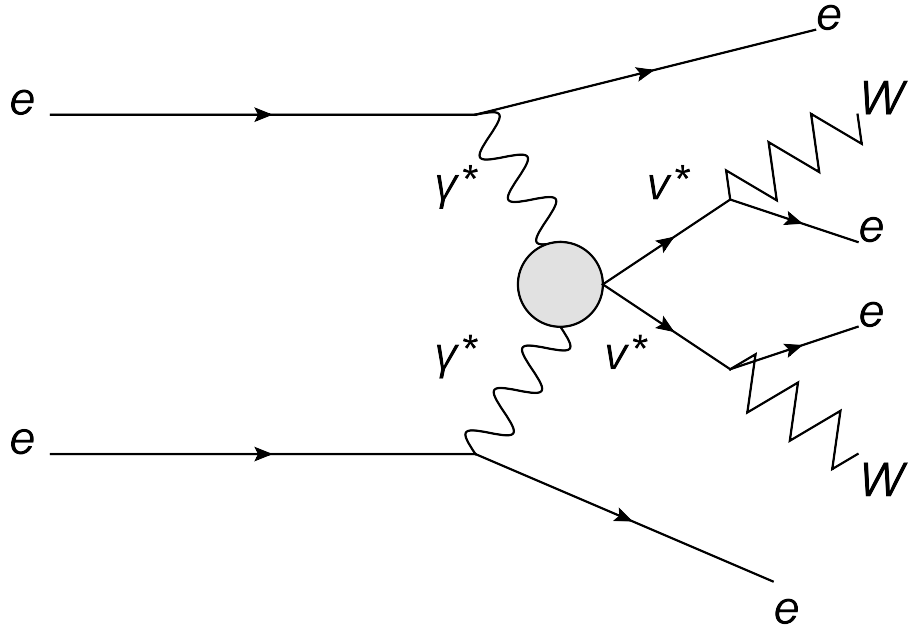}
\caption{Schematic diagram for the process $e^{+}e^{-} \rightarrow e^{+}\gamma^{*}\gamma^{*} e^{-} \rightarrow e^{+}\nu^{*} \bar{\nu}^{*} e^{-}$.
\label{fig8}}
\end{figure}

\begin{figure}
\includegraphics{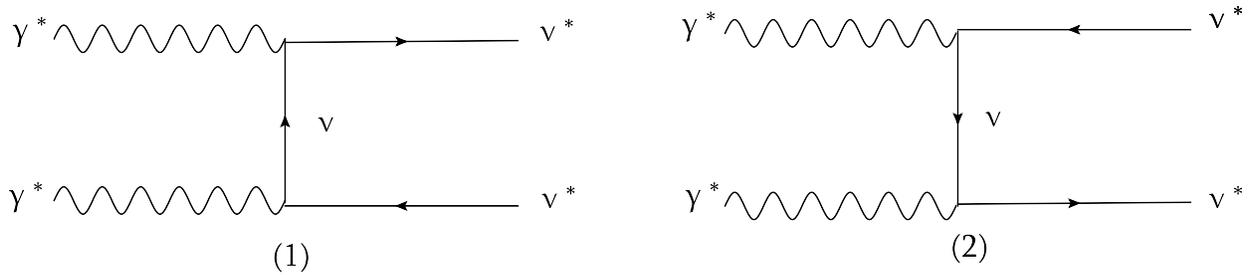}
\caption{Tree-level Feynman diagrams for the subprocess $\gamma^{*}\gamma^{*} \rightarrow \nu^{*} \bar{\nu}^{*}$.
\label{fig5}}
\end{figure}

\begin{figure}
\includegraphics{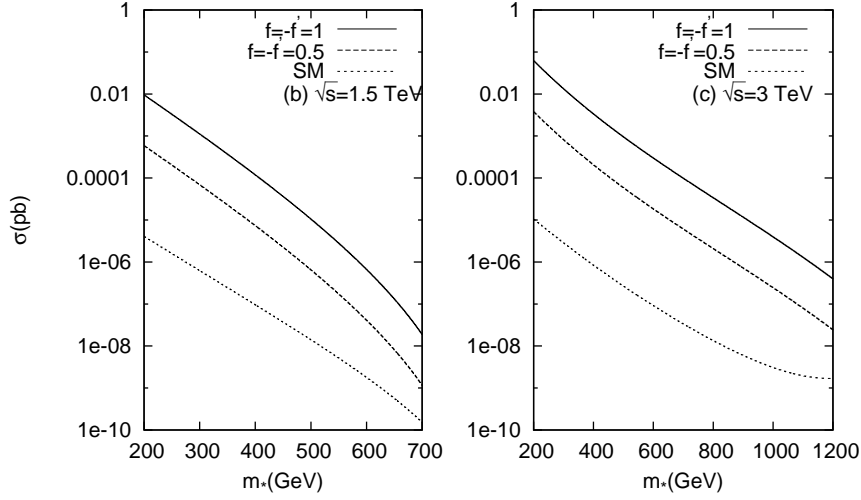}
\caption{The SM and new physics total cross sections of the process $e^{+}e^{-} \rightarrow e^{+}\gamma^{*}\gamma^{*} e^{-} \rightarrow e^{+}\bar{\nu}^{*}\nu^{*} e^{-}\rightarrow e^{+} e^{-} W^{+} e^{+} W^{-} e^{-}$  as a function of excited neutrino mass at various values of coupling parameter and center-of-mass energy.
\label{fig10}}
\end{figure}

\begin{figure}
\includegraphics{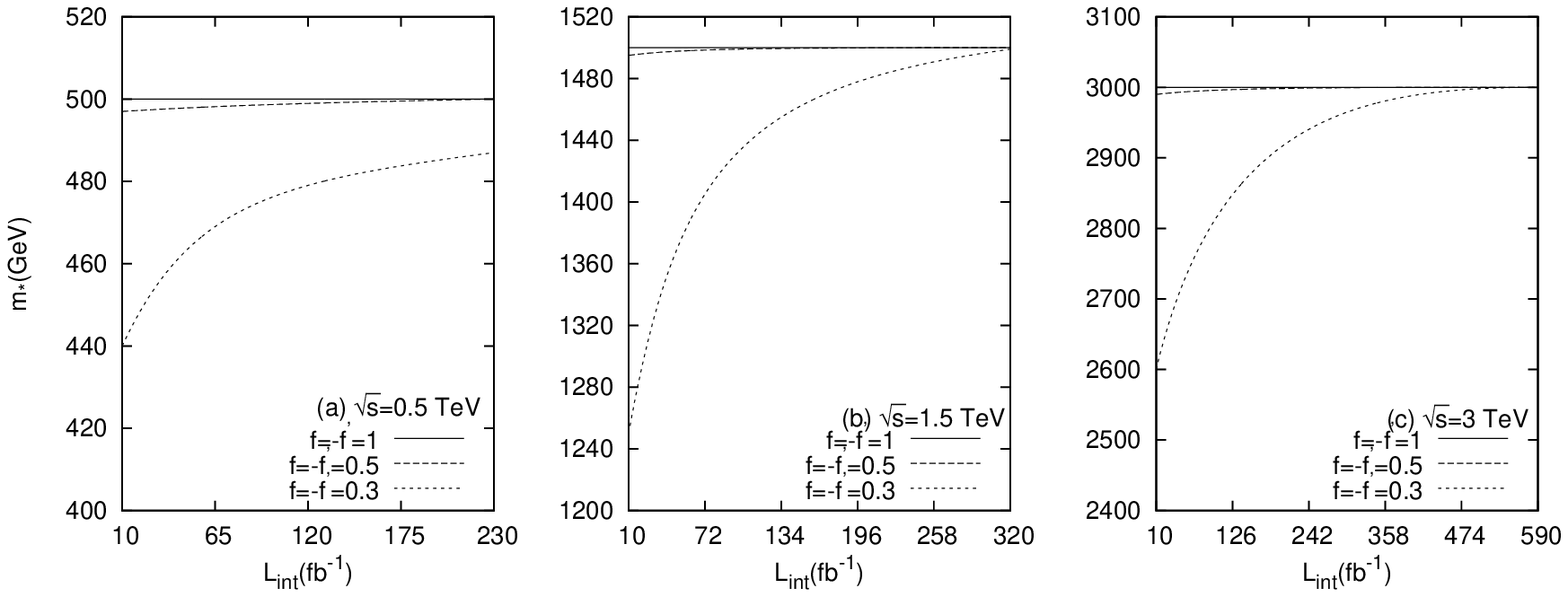}
\caption{$95\%$ C.L. limits for excited neutrino mass
as a function of integrated CLIC luminosity for $\sqrt{s}=0.5,1.5$ and $3$ TeV for the process $e^{+}e^{-}\rightarrow \nu^{*} \bar{\nu}\rightarrow e^{-} W^{+} \bar{\nu}$.
\label{fig11}}
\end{figure}

\begin{figure}
\includegraphics{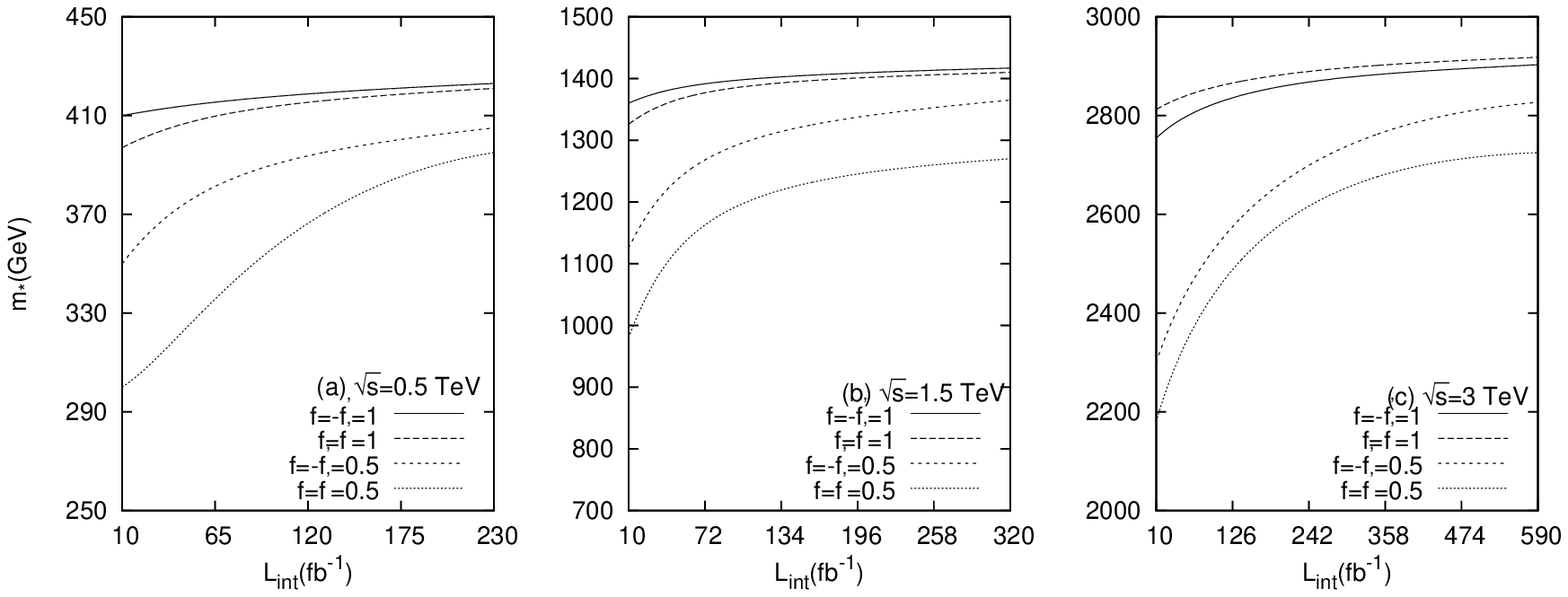}
\caption{$95\%$ C.L. limits for excited neutrino mass
as a function of integrated CLIC luminosity for $\sqrt{s}=0.5,1.5$ and $3$ TeV for the process $e^{+}e^{-} \rightarrow e^{+}\gamma^{*} e^{-} \rightarrow e^{+}\nu^{*} W^{-}\rightarrow e^{+}e^{-}W^{+}W^{-}$.
\label{fig12}}
\end{figure}

\begin{figure}
\includegraphics{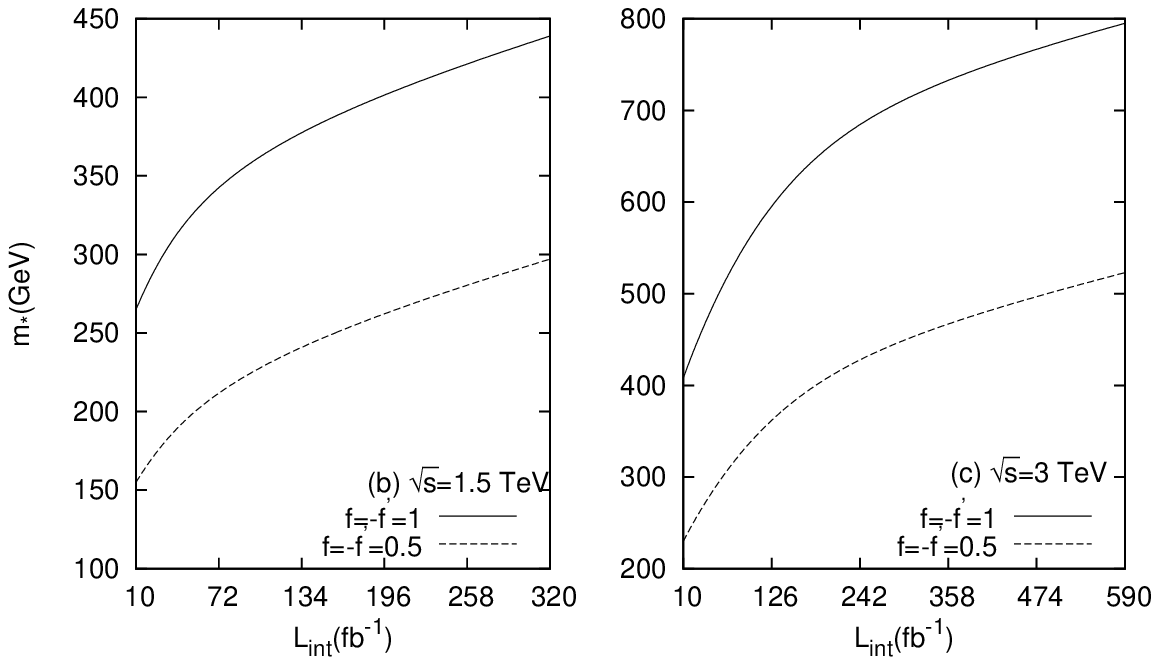}
\caption{$95\%$ C.L. limits for excited neutrino mass
as a function of integrated CLIC luminosity for $\sqrt{s}=0.5,1.5$ and $3$ TeV for the process $e^{+}e^{-} \rightarrow e^{+}\gamma^{*}\gamma^{*} e^{-} \rightarrow e^{+}\bar{\nu}^{*}\nu^{*} e^{-}\rightarrow e^{+} e^{-} W^{+} e^{+} W^{-} e^{-}$.
\label{fig13}}
\end{figure}

\begin{table}
\caption{The branching ratios ($\%$) and total decay widths depending on the mass of excited neutrino for the coupling parameters $f=f^{\prime}=1$ and the scale $\Lambda=m_{*}$ .
\label{tab1}}
\begin{ruledtabular}
\begin{tabular}{cccc}
$m_{*}$(GeV) & $\nu^{*}\rightarrow \nu Z $& $\nu^{*}\rightarrow e W$& $\Gamma_{tot}$(GeV) \\
\hline
$200$& $37$& $63$& $1.02$ \\
$400$& $39$& $61$& $2.61$  \\
$600$& $39$& $61$& $4.06$  \\
$800$& $39$& $61$& $5.49$ \\
$1000$& $39$& $61$& $6.89$  \\
$1200$& $39$& $61$& $8.28$ \\
$1400$& $39$& $61$& $9.70$  \\
$1800$& $39$& $61$& $12.51$  \\
$2200$& $39$& $61$& $15.30$  \\
$2600$& $39$& $61$& $18.03$  \\
\end{tabular}
\end{ruledtabular}
\end{table}

\begin{table}
\caption{The branching ratios ($\%$) and total decay widths depending on the mass of excited neutrino for the coupling parameters $f=-f^{\prime}=1$ and the scale $\Lambda=m_{*}$ .
\label{tab2}}
\begin{ruledtabular}
\begin{tabular}{ccccc}
$m_{*}$(GeV) & $\nu^{*}\rightarrow \nu Z $& $\nu^{*}\rightarrow \nu \gamma $& $\nu^{*}\rightarrow e W$& $\Gamma_{tot}$(GeV) \\
\hline
$200$& $10$& $34$& $56$& $1.14$ \\
$400$& $11$& $29$& $60$& $2.66$  \\
$600$& $11$& $29$& $60$& $4.11$  \\
$800$& $11$& $29$& $60$& $5.51$ \\
$1000$& $11$& $28$& $61$& $6.93$  \\
$1200$& $11$& $28$& $61$& $8.33$ \\
$1400$& $11$& $28$& $61$& $9.73$  \\
$1800$& $11$& $28$& $61$& $12.53$  \\
$2200$& $11$& $28$& $61$& $15.32$  \\
$2600$& $11$& $28$& $61$& $18.05$  \\
\end{tabular}
\end{ruledtabular}
\end{table}

\begin{table}
\caption{Three stages of the CLIC. Here $\sqrt{s}$ is the center-of-mass energy, $N$ is the number of particles in bunch, $L$ is the total luminosity, $\sigma_{x}$
and $\sigma_{y}$ are the beam sizes and $\sigma_{z}$ is the bunch length \cite{18}.
\label{tab1}}
\begin{ruledtabular}
\begin{tabular}{ccccc}
Parameter& Unit& Stage$1$& Stage$2$& Stage$3$ \\
\hline
$\sqrt{s}$& TeV& $0.5$& $1.5$& $3$ \\
$N$& $10^{9}$& $3.7$& $3.7$& $3.7$  \\
$L$& $10^{34}$ cm$^{-2}$ s$^{-1}$& $2.3$& $3.2$& $5.9$  \\
$\sigma_{x}$& nm& $100$& $60$& $40$ \\
$\sigma_{y}$& nm& $2.6$& $1.5$& $1$  \\
$\sigma_{z}$& $\mu$m& $44$& $44$& $44$  \\
\end{tabular}
\end{ruledtabular}
\end{table}

\end{document}